\documentclass[11pt]{article}
\usepackage{jheppub}
\usepackage[utf8]{inputenc}
\usepackage{amsmath}
\usepackage{graphicx}
\usepackage{xcolor}
\usepackage{comment}
\usepackage[toc,page]{appendix}
\usepackage{feynmp}
\usepackage{tikz-feynman}
\tikzfeynmanset{compat=1.1.0}
\usetikzlibrary{positioning,arrows}
\usetikzlibrary{decorations.pathmorphing}
\usetikzlibrary{decorations.markings}

\graphicspath{{Images/}}

\title{Nonrenormalization Theorem \\
for ${\cal N}=(4,4)$  Interface Entropy}
\date{\today}
\author{Andreas Karch${\rm ,}^{1}$}
\author{Hirosi Ooguri${\rm ,}^{2,3}$ and}
\author{Mianqi Wang${}^{1}$}

\affiliation{${}^{1}$Theory Group, Weinberg Institute, Department of Physics,
University of Texas \\  2515 Speedway, Austin, TX 78712, USA.}

\affiliation{${}^{2}$Walter Burke Institute for Theoretical Physics, California Institute of Technology\\ 
 Pasadena, CA 91125, USA
}

\affiliation{${}^{3}$Kavli Institute for the Physics and Mathematics of the Universe (WPI), University of Tokyo \\  
Chiba 277-8583, Japan}

\emailAdd{karcha@utexas.edu}
\emailAdd{ooguri@caltech.edu}
\emailAdd{mqwang@utexas.edu}

\abstract
{We derive a formula for the half-BPS interface entropy between any pair of ${\cal N}=(4,4)$ theories on the same conformal manifold.
This generalizes the diastasis formula derived 
in \cite{Bachas_2014} for ${\cal N}=(2,2)$ theories, which is restricted to the conformal submanifolds generated by either 
chiral or twisted chiral multiples of ${\cal N}=(2,2)$ supersymmetry. 
To derive the ${\cal N}=(4,4)$ formula, we use the fact that the conformal manifold of ${\cal N}=(4,4)$ theories is symmetric and quaternionic-K\"ahler and that its isotropy group contains the $SU(2) \otimes SU(2)$ external automorphism of the ${\cal N}=(4,4)$ superconformal algebra.
As an application of the formula, we prove a supersymmetric non-renormalization theorem, which
explains the observation in \cite{Chiodaroli:2010ur} that the  interface entropy for half-BPS Janus solutions 
in type IIB supergravity on 
${\it AdS}_3 \times S^3 \times T^4$ 
coincides with the corresponding quantity in their free conformal field limits. }

\begin{document}
\tikzset{
line/.style={thick, decorate, draw=black,}
 }
\maketitle
\section{Introduction}

The concept of a {\it space of quantum field theories} plays an important role in the modern study of quantum field theory. For conformal field theories, connected components of the theory space are called conformal manifolds, and many deep and useful results have been discovered about them. Conformal manifolds are Riemannian manifolds, with the Zamolodchikov metric \cite{Zamolodchikov:1986gt} defined in terms of the two-point functions of exactly marginal operators. When the conformal field theory has a weakly coupled holographic dual, the Distance Conjecture \cite{Ooguri:2006in} in anti-de Sitter space (AdS) can be stated in the language of conformal field theory (CFT) \cite{Perlmutter:2020buo}, and parts of it have been proven using CFT techniques \cite{Baume:2023msm, Ooguri:2024ofs}. 

It is desirable to generalize the concept of distance and define it between arbitrary pairs of theories, including those that are not necessarily connected by continuous deformations \cite{Douglas:2010ic}. For conformal field theories,  conformal interfaces offer a promising approach \cite{Oshikawa:1996ww, Oshikawa:1996dj,Bachas:2001vj}. There are several quantities to characterize a conformal interface, such as the energy transmission coefficient \cite{Meineri:2019ycm}
computed by the two-point function of the stress-energy tensor across the interface, the effective central charge \cite{Sakai2008, Brehm2015,Brehm2016}, which controls the amount of entanglement across the interface, and the interface entropy \cite{Affleck:1991tk}. We will focus on the last one in this paper. 

In this paper, we derive a formula for the entropy for the interface preserving one half of the ${\cal N}=(4,4)$ superconformal symmetry 
in two dimensions.
The interface entropy for ${\cal N}=(2,2)$ theories was derived in \cite{Bachas_2014} for a pair of points restricted to the submanifolds 
generated by either chiral or twisted chiral multiples, but not both.
 Our formula generalizes theirs to ${\cal N}=(4,4)$ theories and is applicable to any pair of points on the full conformal manifold. To derive the  formula, we use the fact that the conformal manifold of $N=(4,4)$ theories is symmetric and quaternionic-K\"ahler and that its isotropy group contains the $SU(2) \otimes SU(2)$ external automorphism of the ${\cal N}=(4,4)$ superconformal algebra.

As an application of our formula, we consider the D1-D5 system, where D1 branes are located on points on $T^4$ and D5 branes wrap $T^4$. Its low energy limit 
is described by the supersymmetric sigma-model with the target space
$(T^4)^N/S_N$, where $N$ is the product of the numbers of D1 and D5 branes and $S_N$ is the symmetric group exchanging $N$ copies of $T^4$. The holographic dual is type IIB 
superstring theory on ${\it AdS}_3 \times S^3 \times T^4$.  
In \cite{Chiodaroli:2010ur}, the interface entropy in this setup was computed both in the free CFT limit and in the weakly coupled gravity limit, and  both computations gave the same expression. We will use our formula to explain why they should agree.

This paper is organized as follows. In section 2, we will review the results of \cite{Bachas_2014} that the boundary entropy between ${\cal N}=(2,2)$ theories is given by the K\"ahler potential on the conformal manifold. In section 3, we derive the nonrenormalization theory using the fact that the conformal manifolds of  ${\cal N}=(4,4)$ theories are symmetric spaces. 
In section 4, we will discuss future directions including possible generalization of our results to the transmission coefficient and the effective central charge.

\section{${\cal N}=(2,2)$ and Calabi's Diastasis}

In this section, we review the results of \cite{Bachas_2014} relevant to this paper. Though they apply to any ${\cal N}=(2,2)$ CFTs, it is sometime useful to use the language of the Calabi-Yau sigma-model. Exactly marginal operators of the theory come from
either $(c,c)$ primary fields  or $(c, a)$ primary fields. 
The conformal manifold ${\cal M}$  factorizes locally into two K\"ahler manifolds,\footnote{This statement requires some clarification when the supersymmetry is enhanced to ${\cal N}=(4,4)$, as we will discuss in the next section.}  ${\cal M}_{(c, c)}$  and ${\cal M}_{(c, a)}$,
each generated by $(c,c)$ and $(c, a)$ marginal deformations, respectively. In the Calabi-Yau sigma-model, 
${\cal M}_{(c, c)}$ is the complex structure moduli space and 
${\cal M}_{(c, a)}$ is the complexified K\"ahler moduli space. 
We use the convention where $(c, c)$ primary fields are annihilated by the left-moving supercharge $Q^+$ and 
the right-moving supercharge $\bar Q^+$, while $(c, a)$
primary fields are annihilated by $Q^+$ and $\bar Q^-$. The supercurrents corresponding to the supercharges $Q^\pm$ and $\bar Q^\pm$ are denoted by $G^\pm$ and $\bar G^\pm$, and they carry $\pm 1$ charges with respect to the left and right-moving $U(1)$ currents $J$ and $\bar J$. 

Consider a pair of points on the conformal manifold. 
If there is an interface connecting the theories,
CFT$_1$ and CFT$_2$, at these points while preserving one half of the ${\cal N}=(2,2)$ superconformal symmetry, we can fold the interface to have a boundary condition for the tensor product theory, CFT$_1 \otimes \overline{\rm CFT}_2$,
where $\overline{\rm CFT}_2$ is the parity conjugate of CFT$_2$ \cite{Oshikawa:1996ww, Oshikawa:1996dj,Bachas:2001vj}.
In \cite{Ooguri_1996}, boundary conditions preserving one half of superconformal symmetry are classified into the A-type and B-type.
A boundary state $|A\rangle \rangle $ of the A-type satisfies,
\begin{equation}
    (J -\bar J)|A\rangle \rangle=0, ~~~ (G^\pm -i  \bar G^\mp)|A\rangle \rangle=0, ~~~ e^{i (\phi+\bar \phi)}|A\rangle \rangle=
    e^{i\theta}|A\rangle \rangle,
\end{equation}
where $e^{i \phi}$ and $e^{i \bar \phi}$ are spectral flow operators related to the $U(1)$ currents as
$J = i \partial \phi$, $\bar J = i \bar \partial \bar \phi$ and
$e^{i\theta}$ is some constant phase factor. Similarly, A boundary state $|B\rangle \rangle $ of the B-type satisfies, 
\begin{equation}
    (J +\bar J)|B\rangle \rangle=0, ~~~ (G^\pm -i  \bar G^\pm)|B\rangle \rangle=0, ~~~ e^{i (\phi-\bar \phi)}|B\rangle \rangle=
    e^{i\theta}|B\rangle \rangle.
\end{equation}
In the Calabi-Yau sigma-model, each A-type 
boundary condition is identified to a special Lagrangian submanifold and each B-type boundary condition to a holomorphic submanifold. 

By unfolding, we can represent the interface as an operator that maps the Hilbert space of CFT$_1$ to that of CFT$_2$. If the folded boundary condition is of the A-type, the interface operator $\Delta_A$ satisfies,
\begin{equation}
  [J - \bar J, \Delta_A] = [G^\pm -i \bar G^\mp, \Delta_A] = 0, ~~~
  e^{i(\phi+ \bar \phi)} \Delta_A =
  e^{i\theta}  \Delta_A e^{i(\phi+ \bar \phi)} .
\end{equation}
Similarly, the B-type interface operator $\Delta_B$ satisies,
\begin{equation}
  [J + \bar J, \Delta_B] = [G^\pm -i \bar G^\pm, \Delta_B] = 0, ~~~
  e^{i(\phi- \bar \phi)} \Delta_B =
  e^{i\theta}  \Delta_B e^{i(\phi-\bar \phi)} .
  \label{Btype}
\end{equation}
If CFT$_1 =$ CFT$_2$, we can consider the identity interface, whose interface operator $\Delta$ commutes with all the chiral currents, $J$, $\bar J$, $G^\pm$, $\bar G^\pm$, $\phi$, and $\bar \phi$. Thus, we can regard the identity interface as of both A-type and B-type. In the Calabi-Yau sigma-model, the identity interface can be interpreted as a D-brane wrapping the diagonal subspace of the tensor product of the two copies of the Calabi-Yau manifold. 

The symmetry between the A and B-type is broken if we deform CFT$_2$ away from CFT$_1$. In the Calabi-Yau language, if we deform the complex structure of the Calabi-Yau manifold for CFT$_2$ while keeping that for CFT$_1$ unchanged, the diagonal submanifold is not holomorphic anymore; if it were, the two Calabi-Yau manifolds must have the same complex structure. On the other hand, the diagonal submanifold remains Lagrangian as far as their K\"ahler structures are not deformed.
Therefore, the interface connecting the two CFTs must be of the A-type. Similarly, if we deform the K\"ahler structure for CFT$_2$, it is not possible to impose the Lagrangian condition on the diagonal submanifold, and the interface must become the B-type. 
More generally,  the symmetry 
between the A and B-type is broken by exactly marginal deformation as follows. Consider the identity interface and turn on an exactly marginal deformation generated by a $(c,c)$ primary $\phi$ on CFT$_2$. The first derivative with respect to the $(c,c)$ modulus insert the
operator,
\begin{equation}
    \int_{\cal D} G^- \bar G^- \phi - 2i \int_{\partial \cal D} \phi, 
\end{equation}
where ${\cal D}$ is the $2d$ subspace where CFT$_2$ is defined, and ${\partial \cal D}$ is its boundary. The boundary term is needed in order to preserve one half of the superconformal symmetry \cite{hori2000dbranesmirrorsymmetry}. Since $\phi$ carries the left and right $U(1)$ charges $(+1,+1)$, it does not commute with $J+\bar J$. Therefore, the B-type interface condition \eqref{Btype} cannot be satisfied.

For definiteness, let us 
consider A-type interfaces where only $(c,c)$ moduli are deformed.
The interface entropy $g$ is 
defined by 
\begin{equation}
    g = \lim_{\tau \rightarrow \infty} \frac{\langle 0_1 | e^{-H_1\tau} \Delta e^{-H_2\tau} |0_2\rangle}{\sqrt{\langle 0_1 | e^{-H_1\tau}  |0_1 \rangle \langle 0_2 | e^{-H_2\tau} |0_2\rangle}},
\end{equation}
where $H_a$ and $|0_a\rangle$ are the Hamiltonian and the NS-NS vacuum of CFT$_a$ ($a=1,2$), and $\Delta$ is the interface operator between the two CFTs. We can spectral flow the NS-NS ground states to R-R ground states $|0\rangle_{RR}$ and take the limit $\tau\to\infty$ to obtain the alternative expression \cite{Bachas_2014} 
\begin{equation}
    g^2=\frac{_{RR}\langle \bar{0}_1|\Delta|0_2\rangle_{RR}  ~ _{RR}\langle \bar{0}_2|\Delta^\dagger|0_1\rangle_{RR}}{_{RR}\langle \bar{0}_1|0_1\rangle_{RR} ~  _{RR}\langle \bar{0}_2|0_2\rangle_{RR}} .
    \label{eq:gfunc}
\end{equation}

 By the standard $tt^*$ equations \cite{CECOTTI1991359},
 the denominator of \eqref{eq:gfunc} 
 is $e^{-K(t_1, \bar t_1)-K(t_2, \bar t_2)}$, 
 where $(t_a, \bar t_a)_{a=1,2}$ are the $(c,c)$ moduli of CFT$_a$,
 and $K$ is the K\"ahler potential for them. 
 The numerator is uniquely determined 
to be  $e^{-K(\bar{t}_1,t_2)-K(\bar{t}_2,t_1)}$ 
 by the  $tt^*$ equations and the boundary condition 
 when CFT$_1 =$ CFT$_2$
 \cite{hori2000dbranesmirrorsymmetry}. 
 The expression for the interface entropy is then \cite{Bachas_2014}
\begin{equation}
    2 \log g_{(c,c)}({\rm CFT}_1,{\rm CFT}_2) = K(t_1, \bar t_1) + K(t_2, \bar t_2)
    - K(t_1, \bar t_2) - K(t_2, \bar t_1).
    \label{eq:diastasis}
\end{equation}
This combination of the K\"ahler potential is known as the Calabi diastasis function \cite{Calabi}. The same holds for B-type interfaces, where $(t_a, \bar t_a)$ are complexified K\"ahler moduli.

\section{${\cal N}=(4,4)$ and its External Automorphism}

In this section, we will consider theories with ${\cal N}=(4,4)$ superconformal symmetry. Conformal manifolds of ${\cal N}=(4,4)$ theories
are known to be locally symmetric spaces of the form 
\begin{equation}
   {\cal M}= \frac{SO(4,d)}{SO(4)\times SO(d)},
    \label{symmetric}
\end{equation}
for some $d$  \cite{Seiberg:1988pf, Cecotti:1990kz}. 
In general, this coset is the Teichm\"uller space, and
a conformal manifold is its quotient by a discrete subgroup of $SO(4,d)$. Since 
global structure of a conformal manifold does not play a role in the following, 
we will sometime abuse the terminology and call it a conformal manifold. 

The conformal manifold 
is quaternionic K\"ahler, but it is not 
a K\"ahler manifold for $d > 2$. 
It is in contrast to  ${\cal N}=(2,2)$ theories, whose conformal manifolds are K\"ahler and factorize locally into the product of $(c,c)$ and $(c, a)$ moduli spaces. The difference between ${\cal N}=(2,2)$  and ${\cal N}=(4,4)$ cases was explained in \cite{Gomis:2016sab} as due to some anomalies which appear only if supersymmetry is enhanced beyond ${\cal N}=(2,2)$. On the other hand, 
the $(c,c)$ and $(c,a)$ moduli spaces as submanifolds of ${\cal M}$ are K\"ahler and locally of the form, 
\begin{equation}
{\cal M}_{(c, c)} =\frac{SO(2,d)_{(c,c)}}{SO(2)_{(c,c)}\times SO(d)_{(c,c)} }~~~
{\rm and} ~~~ {\cal M}_{(c, a)} = \frac{SO(2,d)_{(c,a)}}{SO(2)_{(c,a)}\times SO(d)_{(c,a)}},
    \label{maximalkahler}
\end{equation}
where $SO(2,d)_{(c,c)}$ and $SO(2,d)_{(c,a)}$ are different subgroups of $SO(4,d)$ in \eqref{symmetric}.
They are maximal K\"ahler  submanifolds of ${\cal M}$.
In a theory with only ${\cal N}=(2,2)$ superconformal symmetry, 
the formula \eqref{eq:diastasis} for the interface entropy holds only between a pair of points on ${\cal M}_{(c, c)} $ or between a pair of points on 
${\cal M}_{(c, a)}$. With  ${\cal N}=(4,4)$ superconformal symmetry, we can generalize this formula for any pair of points anywhere on the full conformal manifold  \eqref{symmetric} as follows.

The ${\cal N}=(4,4)$ superconformal algebra contains four left-moving supercurrents denoted by $G^\pm$ and $\tilde G^\pm$. 
It also contains the $SU(2)$ Kac-Moody currents $J^\mu$ ($\mu = \pm, 3$), 
which combine the supercurrents with opposite $U(1)$ charges into two 
$SU(2)$ doublets: $(G^-, \tilde G^+)$ and $(\tilde G^-, G^+)$. 
The algebra also has an
external automorphism $SU(2)_{\rm ext}$, which mixes the supercurrents with the same $U(1)$ charges \cite{Berkovits:1994vy, Ooguri:1995cp}. 
It generates a sphere worth of inequivalent
basis of the left-moving supercurrents given by
\begin{equation}
    \begin{aligned}
\tilde G^+(u) &= u^1 \tilde G^+ + u^2 G^+ ,\\
G^-(u) &= u^1  G^- - u^2 \tilde G^-, \\
\tilde G^-(u) &= u^{2*} \tilde G^- - u^{1*} G^- ,\\
G^+(u) &= u^{2*}  G^+ + u^{1*} \tilde G^+ ,\\
    \end{aligned}
    \label{sphereparametrization}
\end{equation}
where 
\begin{equation}
    |u^1|^2 + |u^2|^2 = 1,
\end{equation}
and $u^{A*}$ ($A=1,2$) is the complex conjugate of $u^B\epsilon^{BA}$. There is also a sphere worth of inequivalent basis of the right-moving supercurrents parametrized by $\bar u^A$ and $\bar u^{A*}$.
These two spheres parametrize orbits of the
$SU(2)_{\rm ext} \times SU(2)_{\rm ext}=SO(4)$ external automorphism on the supercurrents.

For a given $(c,c)$ primary field $\phi$, we can construct 
$4$ linearly independent exactly marginal operators. 
Among them,
$G^- \bar G^- \phi$ and $\tilde G^- \bar{\tilde G}^- \phi$ generate $(c,c)$ and $(a,a)$ deformations, 
$G^- \bar{\tilde G}^- \phi$ and $\tilde G^- \bar G^- \phi$ generate $(c,a)$ and $(a,c)$ deformations.
They are mixed by the $SO(4)$ external automorphism and
form its irreducible representation. 
Since exactly marginal deformations correspond to tangent vectors of the conformal manifold, $SO(4)$ acts  as a subgroup of the isotropy group of ${\cal M}$  \cite{Aspinwall:1996mn}. Indeed, the denominator of \eqref{symmetric} shows the isotropy group of ${\cal M}$ is $SO(4) \times SO(d)$, whose $SO(4)$ factor is identified with the external automorphism.

We can combine the $SO(4)$ external automorphism
and the $tt^*$ equations
to derive the coset structure  \eqref{symmetric} of the conformal manifold  \cite{CECOTTI1991359,Boer_2009}. The $tt^*$ equations imply that the Riemann curvature of the conformal manifold can be expressed in terms of the Zamolodchikov metric and the Yukawa couplings -- the structure constant of the chiral ring. The Yukawa couplings for $(c,c)$ primary fields are holomorphic in $(c,c)$ moduli and independent of $(c,a)$ moduli. Combined with the $SO(4)$ automorphism, this implies that the Yukawa couplings are constant on the conformal manifold. Therefore, the Riemann curvature is also covariantly constant. Since the Teichm\"uller space ${\cal M}$ is simply connected and geodesically complete, it follows that it is homogeneous and symmetric. See \cite{CECOTTI1991359,Boer_2009} for more details.

The $SO(4)$ external automorphism can also be used to compute the interface entropy $g$.
Since the $4$ linearly independent marginal operators associated to a given $\phi$ make an irreducible representation of the $SO(4)$ external automorphism, for any pair of points $t_1$ and $t_2$, we can find $\phi$ and $(u, \bar u)$ such that the two points are connected by a marginal deformation generated by a linear combination of $G^-(u) \bar G^-(\bar u) \phi$ and $G^+(u) \bar G^+(u) \phi^*$. 
Since $G^\pm(u)$ and  $\bar G^\pm(\bar u)$ generate an ${\cal N}=(2,2)$ subalgebra of the ${\cal N}=(4,4)$ superconformal algebra, we can treat $G^-(u) \bar G^-(\bar u) \phi$ and $G^+(u) \bar G^+(u) \phi^*$ as $(c,c)$ and $(a,a)$ deformations with respect to this ${\cal N}=(2,2)$ subalgebra.
We can then consider an A-type interface for this  ${\cal N}=(2,2)$ subalgebra between CFTs at $t_1$ and $t_2$. The interface preserves one half of $\tilde G^\pm(u)$ and  $\tilde{\bar G}^\pm(\bar u)$ and we can use the argument in the previous section to show that its interface entropy $g$ is given by \eqref{eq:diastasis}.

\section{Non-renormalization Theorem of Interface Entropy}

In this section, we will explain the coincidence observed in \cite{Chiodaroli:2010ur} between $g$'s in the free CFT limit of the sigma-model with the target space
$(T^4)^N/S_N$ and that computed using the Janus solutions in the weakly coupled supergravity limit of type IIB string on $AdS_3 \times S^3 \times T^4$. 

The conformal manifold of the sigma-model is locally of the form,
\begin{equation}
    {\cal M}_{ (T^4)^N/S_N}
    =  \frac{SO(4,5)}{SO(4)\times SO(5)},
\end{equation}
It contains a $16$-dimensional Narain moduli space of $T^4$,
\begin{equation}
    {\cal M}_{ T^4}
    =  \frac{SO(4,4)}{SO(4)\times SO(4)},
\end{equation}
and its orthogonal $4$-dimensional subspace for the blowup of the $S_N$ orbifold singularity \cite{Larsen:1999uk},
\begin{equation}
    {\cal M}_{ S_N}
    =  \frac{SO(4,1)}{SO(4)},
\end{equation}
which describes the size of the blown-up $S^2$ and the $B$-field on it, and the corresponding complex structure moduli. The CFT can be described as a free orbifold theory when the size of $S^2$ is zero and the $B$-field on it is at the one half period away form zero \cite{Aspinwall:1995zi}. On the other hand, the supergravity description in the holographic dual is a good approximation when the size of $S^2$ is large. 

The conformal manifold ${\cal M}_{ (T^4)^N/S_N}$ also contains the K\"ahler moduli space ${\cal M}_{(c,a)}$ and
the complex structure moduli space ${\cal M}_{(c,c)}$, both of which are homogeneous spaces of the form $SO(2,5)/SO(2) \times SO(5)$. Since the size of the blow-up $S^2$ and the corresponding $B$-field are in the K\"ahler moduli space, the orbifold limit and the supergravity limit are connected by a path on ${\cal M}_{(c,a)}$.

In the supergravity limit,  
the solution describing the half-BPS interface are often referred to as the super-Janus solution as they are supersymetric generalization of the $3d$ Janus construction introduced in \cite{Bak:2007jm}.
In \cite{Chiodaroli:2010ur}, the interface entropy was computed for CFT$_1$ and CFT$_2$ with different K\"ahler moduli on $T^4$, both in the orbifold limit and in the supergravity limit, and both computations gave the same expression for $g$. We can now explain this coincidence. 

Let us denote the conformal manifold coordinates of the 
two CFTs in the orbifold limit with  different K\"ahler moduli on $T^4$ by $(t_1, \bar t_1)$ and $(t_2, \bar t_2)$ in  ${\cal M}_{(c,a)}$, and those of the corresponding CFTs in the supergravity limit by  $(s_1, \bar{s}_1)$ and $(s_2, \bar{s}_2)$. These four points are mapped into each other by the $SO(2,5)$ isometry of  ${\cal M}_{(c,a)}$. In particular, 
$(t_a, \bar{t}_a)$ is mapped to  $(s_a, \bar{s}_a)$  by the same element $\gamma \in SO(2,5)$ for both $a=1$ and $2$. 

Let us compare the formula \eqref{eq:diastasis} in the orbifold limit,
\begin{equation}
    2 \log g_{{\rm orbifold}} = K(t_1, \bar t_1) + K(t_2, \bar t_2)
    - K(t_1, \bar t_2) - K(t_2, \bar t_1),
\end{equation}
and the supergravity limit,
\begin{equation}
    2 \log g_{{\rm sugra}} = K(s_1, \bar{s}_1) + K(s_2, \bar{s}_2)
    - K(s_1, \bar{s}_2) - K(s_2, \bar{s}_1),
\end{equation}
Under an isometry $t_a \rightarrow \gamma\cdot t_a$ by
$\gamma \in SO(2,5)$, the K\"ahler potential transforms as, 
\begin{equation}
    K(\gamma\cdot t_a,\gamma \cdot \bar{t}_a)= K(t_a,\bar{t}_a)+f_\gamma(t_a)+\bar{f}_\gamma (\bar{t}_a)
\end{equation}
for certain holomorphic and anti-holomorphic functions $f_\gamma(t)$ and $\bar{f}_\gamma (\bar{t})$.
Similarly, their analytical continuations transforms by
\begin{equation}
    \begin{aligned}
    K(\gamma \cdot t_1,\gamma \cdot\bar{t}_2)&= K(t_1,\bar{t}_2)+f_\gamma(t_1)+\bar{f}_\gamma (\bar{t}_2)\\
    K(\gamma\cdot t_2,\gamma \cdot \bar{t}_1)&= K(t_2, \bar{t}_1)+f_\gamma(t_2)+\bar{f}_\gamma (\bar{t}_1) . \
\end{aligned}
\end{equation}
In the combination of the K\"ahler potentials in \eqref{eq:diastasis}, the $f_r$ and $\bar{f}_r$ functions cancel out. Hence, we have proved the nonrenormalization theorem of the interface entropy $g$ under the isometry of ${\cal M}_{(c,a)}$:
\begin{equation}
    g(\gamma \cdot t_1, \gamma \cdot\bar t_1; \ \gamma \cdot t_2, \gamma\cdot \bar t_2) = 
    g(t_1, \bar t_1; \ t_2, \bar t_2).
    \label{nonrenormalization}
\end{equation}
Since $(t_a, \bar{t}_a)$ and  $(s_a, \bar{s}_a)$ are mapped into each other by the same elementry of $SO(2,5)$ for both $a=1$ and $2$, the interface entropy computed in the orbifold limit and the supergravity limit should agree. This explains the coincidence found in \cite{Chiodaroli:2010ur}.

\section{Discussion}

We have shown that supersymmetry can be used to prove a nonrenormalization theorem \eqref{nonrenormalization} of the interface entropy $g$ in $2d$ conformal field theories. This nonrenormalization theorem explains the match between field theory and gravity result in the super-Janus interface CFT.
A similar agreement has been observed for two other quantities characterizing supersymmetric interfaces in two-dimensional CFTs with weak gravity dual: the transmission coefficient $c_{LR}$ \cite{baig2024transmissioncoefficientsuperjanussolution}, and the effective central charge associated with entanglement entropy across the interface, $c_{\text{eff}}$ \cite{Gutperle_2016}. The match between CFT and gravity sides for $c_{\text{eff}}$ and $c_{LR}$ also heavily relies on supersymmetry, since the same nonrenormalization fails to hold in the bosonic 3d Janus solution \cite{Bak:2007jm}, which describes a non-supersymmetric interface in the same bulk CFT \cite{Gutperle_2016,Baig_2023}.

A proof of a nonrenormalization theorem based on supersymmetry for $c_{LR}$ and $c_{\text{eff}}$, similar to the one presented here for $g$, remains to be found. $c_{LR}$ is defined in terms of a energy momentum 2-point function \cite{Meineri:2019ycm}, so it is conceivable that techniques similar to the ones employed here can once more be applied. How one would prove non-renormalization of $c_{\text{eff}}$ is more puzzling. One way to approach this problem would be to understand how these quantities transform under the isometry of the conformal manifold, which take us from the weak to strong coupling.

\section*{Acknowledgments}

We thank Paul Aspinwall, Sergio Cecotti, Michael Gutperle, and Ling-Yan Hung 
for discussion. We also thank Costas Bachas, Ilka Brunner, and Michael Gutperle for their comments on the draft of this paper.  
AK and MW are supported in part by the U.S. Department of Energy, Office of High Energy Physics, under Grant No. DE-SC0022021 and a grant from the Simons Foundation (Grant 651678, AK). 
HO is supported in part by the U.S. Department of Energy, Office of Science, Office of High Energy Physics, under Award Number DE-SC0011632, the Simons Investigator Award (MP-SIP-00005259), and JSPS Grants-in-Aid for Scientific Research 23K03379. 
His work was performed in part at the Kavli Institute for the Physics and Mathematics of the Universe at the University of Tokyo, which is supported by the World Premier International Research Center Initiative, MEXT, Japan, and at the Aspen Center for Physics, which is supported
by NSF grant PHY-1607611.

\bibliographystyle{JHEP}
\bibliography{references}

\providecommand{\href}[2]{#2}\begingroup\raggedright\begin{thebibliography}{10}

\bibitem{Bachas_2014}
C.P.~Bachas, I.~Brunner, M.R.~Douglas and L.~Rastelli, \emph{Calabi’s diastasis as interface entropy}, \href{https://doi.org/10.1103/physrevd.90.045004}{\emph{Physical Review D} {\bfseries 90} (2014) }.

\bibitem{Chiodaroli:2010ur}
M.~Chiodaroli, M.~Gutperle and L.-Y.~Hung, \emph{{Boundary entropy of supersymmetric Janus solutions}}, \href{https://doi.org/10.1007/JHEP09(2010)082}{\emph{JHEP} {\bfseries 09} (2010) 082} [\href{https://arxiv.org/abs/1005.4433}{{\ttfamily 1005.4433}}].

\bibitem{Zamolodchikov:1986gt}
A.B.~Zamolodchikov, \emph{{Irreversibility of the Flux of the Renormalization Group in a 2D Field Theory}}, {\emph{JETP Lett.} {\bfseries 43} (1986) 730}.

\bibitem{Ooguri:2006in}
H.~Ooguri and C.~Vafa, \emph{{On the Geometry of the String Landscape and the Swampland}}, \href{https://doi.org/10.1016/j.nuclphysb.2006.10.033}{\emph{Nucl. Phys. B} {\bfseries 766} (2007) 21} [\href{https://arxiv.org/abs/hep-th/0605264}{{\ttfamily hep-th/0605264}}].

\bibitem{Perlmutter:2020buo}
E.~Perlmutter, L.~Rastelli, C.~Vafa and I.~Valenzuela, \emph{{A CFT distance conjecture}}, \href{https://doi.org/10.1007/JHEP10(2021)070}{\emph{JHEP} {\bfseries 10} (2021) 070} [\href{https://arxiv.org/abs/2011.10040}{{\ttfamily 2011.10040}}].

\bibitem{Baume:2023msm}
F.~Baume and J.~Calder\'on-Infante, \emph{{On higher-spin points and infinite distances in conformal manifolds}}, \href{https://doi.org/10.1007/JHEP12(2023)163}{\emph{JHEP} {\bfseries 12} (2023) 163} [\href{https://arxiv.org/abs/2305.05693}{{\ttfamily 2305.05693}}].

\bibitem{Ooguri:2024ofs}
H.~Ooguri and Y.~Wang, \emph{{Universal Bounds on CFT Distance Conjecture}},  \href{https://arxiv.org/abs/2405.00674}{{\ttfamily 2405.00674}}.

\bibitem{Douglas:2010ic}
M.R.~Douglas, \emph{{Spaces of Quantum Field Theories}}, \href{https://doi.org/10.1088/1742-6596/462/1/012011}{\emph{J. Phys. Conf. Ser.} {\bfseries 462} (2013) 012011} [\href{https://arxiv.org/abs/1005.2779}{{\ttfamily 1005.2779}}].

\bibitem{Oshikawa:1996ww}
M.~Oshikawa and I.~Affleck, \emph{{Defect lines in the Ising model and boundary states on orbifolds}}, \href{https://doi.org/10.1103/PhysRevLett.77.2604}{\emph{Phys. Rev. Lett.} {\bfseries 77} (1996) 2604} [\href{https://arxiv.org/abs/hep-th/9606177}{{\ttfamily hep-th/9606177}}].

\bibitem{Oshikawa:1996dj}
M.~Oshikawa and I.~Affleck, \emph{{Boundary conformal field theory approach to the critical two-dimensional Ising model with a defect line}}, \href{https://doi.org/10.1016/S0550-3213(97)00219-8}{\emph{Nucl. Phys. B} {\bfseries 495} (1997) 533} [\href{https://arxiv.org/abs/cond-mat/9612187}{{\ttfamily cond-mat/9612187}}].

\bibitem{Bachas:2001vj}
C.~Bachas, J.~de~Boer, R.~Dijkgraaf and H.~Ooguri, \emph{{Permeable conformal walls and holography}}, \href{https://doi.org/10.1088/1126-6708/2002/06/027}{\emph{JHEP} {\bfseries 06} (2002) 027} [\href{https://arxiv.org/abs/hep-th/0111210}{{\ttfamily hep-th/0111210}}].

\bibitem{Meineri:2019ycm}
M.~Meineri, J.~Penedones and A.~Rousset, \emph{{Colliders and conformal interfaces}}, \href{https://doi.org/10.1007/JHEP02(2020)138}{\emph{JHEP} {\bfseries 02} (2020) 138} [\href{https://arxiv.org/abs/1904.10974}{{\ttfamily 1904.10974}}].

\bibitem{Sakai2008}
K.~Sakai and Y.~Satoh, \emph{Entanglement through conformal interfaces}, \href{https://doi.org/10.1088/1126-6708/2008/12/001}{\emph{JHEP} {\bfseries 12} (2008) 001} [\href{https://arxiv.org/abs/0809.4548}{{\ttfamily 0809.4548}}].

\bibitem{Brehm2015}
E.M.~Brehm and I.~Brunner, \emph{{Entanglement entropy through conformal interfaces in the 2D Ising model}}, \href{https://doi.org/10.1007/JHEP09(2015)080}{\emph{JHEP} {\bfseries 09} (2015) 080} [\href{https://arxiv.org/abs/1505.02647}{{\ttfamily 1505.02647}}].

\bibitem{Brehm2016}
E.M.~Brehm, I.~Brunner, D.~Jaud and C.~Schmidt-Colinet, \emph{Entanglement and topological interfaces}, \href{https://doi.org/10.1002/prop.201600024}{\emph{Fortsch. Phys.} {\bfseries 64} (2016) 516} [\href{https://arxiv.org/abs/1512.05945}{{\ttfamily 1512.05945}}].

\bibitem{Affleck:1991tk}
I.~Affleck and A.W.W.~Ludwig, \emph{{Universal noninteger 'ground state degeneracy' in critical quantum systems}}, \href{https://doi.org/10.1103/PhysRevLett.67.161}{\emph{Phys. Rev. Lett.} {\bfseries 67} (1991) 161}.

\bibitem{Ooguri_1996}
H.~Ooguri, Y.~Oz and Z.~Yin, \emph{D-branes on calabi-yau spaces and their mirrors}, \href{https://doi.org/10.1016/0550-3213(96)00379-3}{\emph{Nuclear Physics B} {\bfseries 477} (1996) 407–430}.

\bibitem{hori2000dbranesmirrorsymmetry}
K.~Hori, A.~Iqbal and C.~Vafa, \emph{D-branes and mirror symmetry},  2000.

\bibitem{CECOTTI1991359}
S.~Cecotti and C.~Vafa, \emph{Topologicalâ€”anti-topological fusion}, \href{https://doi.org/https://doi.org/10.1016/0550-3213(91)90021-O}{\emph{Nuclear Physics B} {\bfseries 367} (1991) 359}.

\bibitem{Calabi}
E.~Calabi, \emph{Isometric imbedding of complex manifolds}, {\emph{Ann. Math.} {\bfseries 58} (1953) 1}.

\bibitem{Seiberg:1988pf}
N.~Seiberg, \emph{{Observations on the Moduli Space of Superconformal Field Theories}}, \href{https://doi.org/10.1016/0550-3213(88)90183-6}{\emph{Nucl. Phys. B} {\bfseries 303} (1988) 286}.

\bibitem{Cecotti:1990kz}
S.~Cecotti, \emph{{N=2 Landau-Ginzburg versus Calabi-Yau sigma models: Nonperturbative aspects}}, \href{https://doi.org/10.1142/S0217751X91000939}{\emph{Int. J. Mod. Phys. A} {\bfseries 6} (1991) 1749}.

\bibitem{Gomis:2016sab}
J.~Gomis, Z.~Komargodski, H.~Ooguri, N.~Seiberg and Y.~Wang, \emph{{Shortening Anomalies in Supersymmetric Theories}}, \href{https://doi.org/10.1007/JHEP01(2017)067}{\emph{JHEP} {\bfseries 01} (2017) 067} [\href{https://arxiv.org/abs/1611.03101}{{\ttfamily 1611.03101}}].

\bibitem{Berkovits:1994vy}
N.~Berkovits and C.~Vafa, \emph{{N=4 topological strings}}, \href{https://doi.org/10.1016/0550-3213(94)00419-F}{\emph{Nucl. Phys. B} {\bfseries 433} (1995) 123} [\href{https://arxiv.org/abs/hep-th/9407190}{{\ttfamily hep-th/9407190}}].

\bibitem{Ooguri:1995cp}
H.~Ooguri and C.~Vafa, \emph{{All loop N=2 string amplitudes}}, \href{https://doi.org/10.1016/0550-3213(95)00365-Y}{\emph{Nucl. Phys. B} {\bfseries 451} (1995) 121} [\href{https://arxiv.org/abs/hep-th/9505183}{{\ttfamily hep-th/9505183}}].

\bibitem{Aspinwall:1996mn}
P.S.~Aspinwall, \emph{{K3 surfaces and string duality}},  in \emph{{Theoretical Advanced Study Institute in Elementary Particle Physics (TASI 96): Fields, Strings, and Duality}}, pp.~421--540, 11, 1996 [\href{https://arxiv.org/abs/hep-th/9611137}{{\ttfamily hep-th/9611137}}].

\bibitem{Boer_2009}
J.d.~Boer, J.~Manschot, K.~Papadodimas and E.~Verlinde, \emph{The chiral ring of ads3/cft2and the attractor mechanism}, \href{https://doi.org/10.1088/1126-6708/2009/03/030}{\emph{Journal of High Energy Physics} {\bfseries 2009} (2009) 030–030}.

\bibitem{Larsen:1999uk}
F.~Larsen and E.J.~Martinec, \emph{{U(1) charges and moduli in the D1 - D5 system}}, \href{https://doi.org/10.1088/1126-6708/1999/06/019}{\emph{JHEP} {\bfseries 06} (1999) 019} [\href{https://arxiv.org/abs/hep-th/9905064}{{\ttfamily hep-th/9905064}}].

\bibitem{Aspinwall:1995zi}
P.S.~Aspinwall, \emph{{Enhanced gauge symmetries and K3 surfaces}}, \href{https://doi.org/10.1016/0370-2693(95)00957-M}{\emph{Phys. Lett. B} {\bfseries 357} (1995) 329} [\href{https://arxiv.org/abs/hep-th/9507012}{{\ttfamily hep-th/9507012}}].

\bibitem{Bak:2007jm}
D.~Bak, M.~Gutperle and S.~Hirano, \emph{{Three dimensional Janus and time-dependent black holes}}, \href{https://doi.org/10.1088/1126-6708/2007/02/068}{\emph{JHEP} {\bfseries 02} (2007) 068} [\href{https://arxiv.org/abs/hep-th/0701108}{{\ttfamily hep-th/0701108}}].

\bibitem{baig2024transmissioncoefficientsuperjanussolution}
S.~Baig, A.~Karch and M.~Wang, \emph{Transmission coefficient of super-janus solution},  2024.

\bibitem{Gutperle_2016}
M.~Gutperle and J.D.~Miller, \emph{Entanglement entropy at holographic interfaces}, \href{https://doi.org/10.1103/physrevd.93.026006}{\emph{Physical Review D} {\bfseries 93} (2016) }.

\bibitem{Baig_2023}
S.A.~Baig and S.~Shashi, \emph{Transport across interfaces in symmetric orbifolds}, \href{https://doi.org/10.1007/jhep10(2023)168}{\emph{Journal of High Energy Physics} {\bfseries 2023} (2023) }.

\end{thebibliography}\endgroup
\end{document}